\documentclass[12pt,epsf]{article}
\usepackage{graphicx,amsmath,amssymb}
\begin{document}

\def\a{\alpha}
\def\b{\beta}
\def\c{\varepsilon}
\def\d{\delta}
\def\e{\epsilon}
\def\f{\phi}
\def\g{\gamma}
\def\h{\theta}
\def\k{\kappa}
\def\l{\lambda}
\def\m{\mu}
\def\n{\nu}
\def\p{\psi}
\def\q{\partial}
\def\r{\rho}
\def\s{\sigma}
\def\t{\tau}
\def\u{\upsilon}
\def\v{\varphi}
\def\w{\omega}
\def\x{\xi}
\def\y{\eta}
\def\z{\zeta}
\def\D{\Delta}
\def\G{\Gamma}
\def\H{\Theta}
\def\L{\Lambda}
\def\F{\Phi}
\def\P{\Psi}
\def\S{\Sigma}

\def\o{\over}
\def\beq{\begin{eqnarray}}
\def\eeq{\end{eqnarray}}
\newcommand{\gsim}{ \mathop{}_{\textstyle \sim}^{\textstyle >} }
\newcommand{\lsim}{ \mathop{}_{\textstyle \sim}^{\textstyle <} }
\newcommand{\vev}[1]{ \left\langle {#1} \right\rangle }
\newcommand{\bra}[1]{ \langle {#1} | }
\newcommand{\ket}[1]{ | {#1} \rangle }
\newcommand{\EV}{ {\rm eV} }
\newcommand{\KEV}{ {\rm keV} }
\newcommand{\MEV}{ {\rm MeV} }
\newcommand{\GEV}{ {\rm GeV} }
\newcommand{\TEV}{ {\rm TeV} }
\def\diag{\mathop{\rm diag}\nolimits}
\def\Spin{\mathop{\rm Spin}}
\def\SO{\mathop{\rm SO}}
\def\O{\mathop{\rm O}}
\def\SU{\mathop{\rm SU}}
\def\U{\mathop{\rm U}}
\def\Sp{\mathop{\rm Sp}}
\def\SL{\mathop{\rm SL}}
\def\tr{\mathop{\rm tr}}

\def\IJMP{Int.~J.~Mod.~Phys. }
\def\MPL{Mod.~Phys.~Lett. }
\def\NP{Nucl.~Phys. }
\def\PL{Phys.~Lett. }
\def\PR{Phys.~Rev. }
\def\PRL{Phys.~Rev.~Lett. }
\def\PTP{Prog.~Theor.~Phys. }
\def\ZP{Z.~Phys. }

\newcommand{\bea}{\begin{eqnarray}}   
\newcommand{\eea}{\end{eqnarray}}
\newcommand{\bear}{\begin{array}}  
\newcommand {\eear}{\end{array}}
\newcommand{\bef}{\begin{figure}}  
\newcommand {\eef}{\end{figure}}
\newcommand{\bec}{\begin{center}}  
\newcommand {\eec}{\end{center}}
\newcommand{\non}{\nonumber}  
\newcommand {\eqn}[1]{\beq {#1}\eeq}
\newcommand{\la}{\left\langle}  
\newcommand{\ra}{\right\rangle}
\newcommand{\ds}{\displaystyle}
\def\SEC#1{Sec.~\ref{#1}}
\def\FIG#1{Fig.~\ref{#1}}
\def\EQ#1{Eq.~(\ref{#1})}
\def\EQS#1{Eqs.~(\ref{#1})}
\def\TEV#1{10^{#1}{\rm\,TeV}}
\def\GEV#1{10^{#1}{\rm\,GeV}}
\def\MEV#1{10^{#1}{\rm\,MeV}}
\def\KEV#1{10^{#1}{\rm\,keV}}
\def\lrf#1#2{ \left(\frac{#1}{#2}\right)}
\def\lrfp#1#2#3{ \left(\frac{#1}{#2} \right)^{#3}}
\def\REF#1{Ref.~\cite{#1}}
\newcommand{\osc}{{\rm osc}}
\newcommand{\ed}{{\rm end}}
\def\dda#1{\frac{\partial}{\partial a_{#1}}}
\def\ddat#1{\frac{\partial^2}{\partial a_{#1}^2}}
\def\dd#1#2{\frac{\partial #1}{\partial #2}}
\def\ddt#1#2{\frac{\partial^2 #1}{\partial #2^2}}
\def\lrp#1#2{\left( #1 \right)^{#2}}


\baselineskip 0.7cm

\begin{titlepage}

\begin{flushright}
TU-961\\
IPMU14-0085\\
\end{flushright}

\vskip 1.35cm
\begin{center}
{\large \bf 
Anomaly-free flavor models for Nambu-Goldstone bosons\\
and the $3.5$ keV X-ray line signal
}
\vskip 1.2cm
Kazunori Nakayama$^{a,c}$,
Fuminobu Takahashi$^{b,c}$
and 
Tsutomu T. Yanagida$^{c}$

\vskip 0.4cm

{\it $^a$Department of Physics, University of Tokyo, Tokyo 113-0033, Japan}\\
{\it $^b$Department of Physics, Tohoku University, Sendai 980-8578, Japan}\\
{\it $^c$Kavli Institute for the Physics and Mathematics of the Universe (WPI), TODIAS, University of Tokyo, Kashiwa 277-8583, Japan}

\vskip 1.5cm

\abstract{
We pursue a possibility that a pseudo-Nambu-Goldstone boson is lurking around or below the intermediate scale.
To this end we consider an anomaly-free global flavor symmetry, and construct models where the pseudo-Nambu-Goldstone boson is
 coupled preferentially  to leptons. The experimental and astrophysical bounds derived from  couplings to photons and  nucleons are significantly relaxed. 
 If sufficiently light, the pseudo-Nambu-Goldstone boson contributes to dark matter, and interestingly, 
it generally decays into photons through couplings arising from threshold corrections. We show that
the recent hint for the  X-ray line at about $3.5$ keV can be explained by the decay of such pseudo-Nambu-Goldstone boson
of mass about $7$ keV with the decay constant of order $10^{10}$\,GeV, if the electron
is charged  under the flavor symmetry.  
}
\end{center}
\end{titlepage}

\setcounter{page}{2}

\section{Introduction}

Symmetry plays an important role in physics. 
 Sometimes it is spontaneously broken in the low energy, and as a remnant, 
  there appears a massless Nambu-Goldstone boson. 
  If the symmetry is a local one,  it
is absorbed by the corresponding gauge boson. On the other hand, 
if the symmetry is a global and approximate one, there remains 
a light pseudo-Nambu-Goldstone boson (PNGB),  which has been a subject of considerable
interest. 
PNGBs, if exist, will provide us with invaluable information on the high energy physics. 

Various types of global symmetries and the associated PNGBs
have been considered so far. One example  is the QCD axion, which arises in 
association with the spontaneous breakdown of the Peccei-Quinn symmetry~\cite{Peccei:1977hh,QCD-axion}. 
Importantly, the QCD axion is coupled to gluons and photons through anomalies, 
as well as to the quarks and the leptons at tree or one-loop level.  The interactions are suppressed by the decay constant, which parametrizes the symmetry breaking scale.
In other extensions of the standard model (SM), 
there arise PNGBs with similar properties, the so called axion-like particles, and especially those
with couplings to photons have been studied extensively from both theoretical and 
experimental aspects~\cite{Jaeckel:2010ni}.

The couplings of the QCD axion and the axion-like particles to photons, nucleons, and electrons 
are tightly constrained by cosmology, astrophysics and the ground-based 
experiments~\cite{Jaeckel:2010ni,Andreas:2010ms,Cadamuro:2011fd}.
In particular, the astrophysical constraints  are extremely tight, pushing
the scale of new physics to an intermediate scale or above.
Still, there may be other kind of PNGBs with different properties
 at a scale around or even below the intermediate scale,  without any conflict with those constraints.

In this paper we pursue a possibility that a PNGB associated with
new physics is lurking around or below the intermediate scale. For this, we need to evade 
tight  astrophysical bounds on the PNGBs. One way is to consider PNGBs, which are
not directly coupled to the SM sector, but  mainly coupled to a hidden sector~\cite{Weinberg:2013kea}.  
Instead,  we want to consider here the case in which some of the SM particles are charged under
a global flavor symmetry. The maximal possible flavor symmetry for the SM particles with  three right-handed neutrinos 
is ${\rm U(3)}^6$. 
We consider an anomaly-free global U(1)$_F$ flavor symmetry, 
which is a subgroup of the maximal flavor symmetry. In particular, a leptophilic PNGB model
 is simple and phenomenologically interesting, and 
we will construct concrete models along this lines. Such leptophilic PNGBs without anomalous
couplings to photons evade various  experimental and astrophysical bounds coming 
from couplings with  nucleons and photons. We will mainly focus on very light PNGBs with mass
lighter than the twice the electron mass.\footnote{ 
Experimental bounds on PNGBs with mass heavier than ${\cal O}(1)$\,MeV
 including leptophilic ones  were studied in Ref.~\cite{Essig:2010gu}.  
}

There is an interesting point of the PNGB associated with an anomaly-free global symmetry.
Although suppressed, such PNGB is necessarily coupled to photons through
threshold corrections. In particular,  the decay into two photons can be the main decay mode
if the PNGB of mass is  less than twice the electron mass. 
If such light PNGB constitutes dark matter, it mainly decays into two photons,
producing a narrow X-ray line. This can  explain the recent hint for the 
X-ray line at about $3.5$ keV~\cite{Bulbul:2014sua,Boyarsky:2014jta}
for the PNGB mass of about $7$\,keV. As we shall see, the required decay constant is
$f_a = {\cal O}(10^{10})$\,GeV if the electron is charged under the symmetry, whereas
it is $f_a = {\cal O}(10^{5})$\,GeV if the electron is neutral under the symmetry.
This should be contrasted to the fact that the observed X-ray flux can also be explained by the string axion
with a decay constant of order $\GEV{14-15}$ as first pointed out in Ref.~\cite{Higaki:2014zua}.\footnote{
The X-ray line produced by light modulus decay was studied many years ago by Kawasaki and one of the present authors (TTY)
in Ref.~\cite{Kawasaki:1997ah} (see also Refs.~\cite{Hashiba:1997rp,Kusenko:2012ch}). Recently there appeared various 
possibilities to explain the $3.5$\,keV X-ray line~\cite{Ishida:2014dlp,Higaki:2014zua, Finkbeiner:2014sja, Jaeckel:2014qea,Lee:2014xua,Krall:2014dba}. 
}

The rest of the paper is organized as follows. In Sec.~\ref{sec:2} we discuss the coupling of the PNGB
to photons through threshold corrections, and its implications for the $3.5$\,keV X-ray line.
We discuss production of PNGB dark matter in Sec.~\ref{sec:3}. In Sec.~\ref{sec:4}, we
will build concrete models for leptophilic PNGBs.  The last section is devoted for
discussion and conclusions.

\section{Couplings of PNGBs to photons}
\label{sec:2}

Let us consider a global U(1)$_F$ flavor symmetry under which only leptons are charged.
Most important, we assume that the global U(1)$_F$ symmetry is anomaly free so that
the PNGB coupling  to photons is suppressed, evading various observational constraints. 
The coupling to photons is nevertheless induced by threshold corrections,
which we will study in this section.

Let us study the interactions of the PNGB with leptons in the low energy.
Later we will construct concrete flavor models. 
The relevant low-energy interactions are given by
\bea
\label{NGlepton}
- {\cal L} &=&   m_e {\bar e}_R e_L e^{i q_e \frac{a}{f_a}}+ 
m_\mu {\bar \mu}_R \mu_L e^{i q_\mu \frac{a}{f_a}}+ 
m_\tau {\bar \tau}_R \tau_L e^{i q_\tau \frac{a}{f_a}} + {\rm h.c.}
\eea
where $a$ is the PNGB associated  with the flavor symmetry, $f_a$ the decay constant, 
and $q_{e}$, $q_\mu$, and $q_{\tau}$ the coupling constants for electron, muon and tau leptons, respectively. 
We exclude the case of $q_e =q_\mu = q_\tau = 0$ in the following analysis. 

We are interested in the case where the PNGB mass is much lighter than twice the electron mass.
Integrating out electron, muon, and tau leptons, therefore, we obtain the effective interaction,
\bea
{\cal L}_{\rm eff} &\simeq&
-  (q_e+q_\mu+q_\tau) \frac{  \alpha_{em} }{4 \pi f_a} a F_{\mu \nu} {\tilde F}^{\mu \nu}\non\\&&
+ \frac{ \alpha_{em} }{48 \pi f_a} \left( \frac{q_e}{m_e^2} + \frac{q_\mu}{m_\mu^2} + \frac{q_\tau}{m_\tau^2} \right) \left(
(\partial^2 a) F_{\mu \nu} {\tilde F}^{\mu \nu} + 2a  F_{\mu \nu} \partial^2{\tilde F}^{\mu \nu} \right),
\label{Leff1}
\eea
where the first line corresponds to the anomaly term, and the second line arises from the threshold corrections. 
We require $q_e + q_\mu+q_\tau = 0$ to ensure that the flavor symmetry is anomaly-free. 
Then the first term in \EQ{Leff1} vanishes, and we are left with the finite threshold corrections.
Therefore the PNGB coupling to photons  is significantly suppressed for anomaly-free symmetry. 
As long as we are interested in the decay or production of the on-shell PNGB and photons, 
we can use their equations of motion. Then the effective interaction for the PNGB to photons becomes
\bea
{\cal L}_{\rm eff} &=& \frac{ \alpha_{em} m_a^2 }{48 \pi f_a} \left(\frac{q_e}{m_e^2} + \frac{q_\mu}{m_\mu^2} + \frac{q_\tau}{m_\tau^2}\right) a  F_{\mu \nu} {\tilde F}^{\mu \nu}
\eea
for the on-shell PNGB and photons and $m_a^2 \ll m_e^2$. 
The PNGB coupling to photons is dominated by  the first term if $q_e \ne 0$; otherwise it is dominated
by the second term.  Note that both $q_e$ and $q_\mu$ cannot vanish simultaneously to satisfy the anomaly-free condition.

The decay rate of the PNGB  into two photons is approximately given by
\bea
\Gamma_{a \to \gamma \gamma} &\simeq&
 \frac{\alpha_{em}^2}{9216\pi^3} \frac{m_a^7}{f_a^2}  \times
\left\{
\bear{cc}
q_e^2 /m_e^4&~~~{\rm for~~}q_e \ne 0,\\
&\\
q_\mu^2 /m_\mu^4&~~~{\rm for~~}q_e = 0
 \eear
 \right.
\eea
where we have approximated $m_e^2\ll m_\mu^2 \ll m_\tau^2$ and assumed that there is no
large hierarchy among the U(1)$_F$  charges.
Assuming that the PNGB decays mainly into photons via the above interaction, we can 
estimate the lifetime as
\bea
\tau_{a \to \gamma \gamma} &\simeq& 
\left\{
\bear{ll}
\ds{
2.9 \times 10^{28}\, q_e^{-2} \lrfp{m_a}{7 {\rm keV}}{-7} \lrfp{f_a}{\GEV{10}}{2}  {\rm sec.
}}&~~~{\rm for~~}q_e \ne 0\\
\ds{2.1 \times 10^{28}\, q_\mu^{-2} \lrfp{m_a}{7 {\rm keV}}{-7} \lrfp{f_a}{2 \times \GEV{5}}{2}  {\rm sec.}
}&~~~{\rm for~~}q_e = 0
\eear
\right.
\eea
Thus the PNGB is so long-lived that it can contribute to dark matter. 
We will show in the next section that, in fact,  the right amount of PNGBs can be produced to explain the observed
 dark matter abundance. 

The recent hint for the X-ray line at about $3.5$\,keV can be explained by dark matter with the following 
mass and lifetime~\cite{Bulbul:2014sua,Boyarsky:2014jta}:
\bea
m_{\rm DM} &\simeq&7.1 {\rm \, keV},\\
\tau_{\rm DM} &\simeq& 4 \times 10^{27}  - 4 \times 10^{28}\, {\rm sec},
\label{tau-obs}
\eea
if it decays into a pair of photons. 
Therefore, the $3.5$\,keV X-ray line  can be explained by the decay of the PNGB dark matter
with  $m_a \simeq 7$\,keV and $f_a/q_e = 4\times \GEV{9} - 1\times \GEV{10} $ for $q_e \ne 0$,  
or $f_a/q_\mu =  9 \times \GEV{4} - 3\times\GEV{5} $ for $q_e = 0$ and $q_\mu \ne 0$.
 Interestingly, 
a relatively small decay constant below the intermediate scale is needed  because of  the suppression factor
for  the threshold corrections.
This should be contrasted to the fact that the observed X-ray flux can also be explained by the string axion
with a decay constant of order $\GEV{14-15}$~\cite{Higaki:2014zua}.

\section{PNGB dark matter}
\label{sec:3}

A light PNGB contributes to dark matter, if it is sufficiently long-lived. In order to explain the observed dark matter density,
 the right amount of PNGBs need to  be produced in the early 
Universe. There are two important production processes. One is 
non-thermal production by the initial misalignment mechanism, and the other is
thermal production.\footnote{The production of PNGB dark 
matter was recently studied in Ref.~\cite{Jaeckel:2013uva}.}
We will consider these production processes in turn.

The PNGB number density to entropy ratio  can be written  as
\bea
Y_a &\simeq& 6 \times 10^{-5} \lrfp{m_a}{7 {\rm keV}}{-1} \lrf{\Omega_a h^2}{0.12},
\eea
where $\Omega_a$ is the density parameter for the PNGB and $h$ is the reduced Hubble constant. 
On the other hand, if the PNGB is in equilibrium, its abundance is given by
\bea
Y_a^{\rm (eq)} &\simeq&2.6 \times 10^{-3} \lrfp{g_*}{106.75}{-1},
\eea
where $g_*$ counts the relativistic degrees of freedom in thermal plasma. 
Therefore, if the PNGBs constitute the observed dark matter, they should not be in equilibrium,
otherwise there must be late-time entropy dilution by a factor $40$ for $m_a \simeq 7$keV.

Let us first consider the case of $q_e \ne 0$. In this case, the decay constant suggested by the
observed X-ray line is $f_a/q_e = 4\times \GEV{9} - 1\times \GEV{10}$. The thermal production 
process depends on the charge of $\tau$. If the PNGB is directly coupled to $\tau$, the main production
process will be through scatterings between leptons and Higgs bosons such as $\ell_3 H^* \to a \tau_R$.
The abundance is roughly estimated as follows
\bea
Y_a^{{\rm (th)}} &\sim&6 \times 10^{-5} \lrfp{g_*}{106.75}{-1} \lrf{T_R}{\GEV{6}} \lrfp{f_a}{\GEV{10}}{-2},
\eea
where $T_R$ is the reheating temperature. Thus, the right amount of PNGBs are thermally
produced for $T_R \sim \GEV{6}$ and $f_a \sim \GEV{10}$.
Alternatively, if the PNGB is not directly coupled to $\tau$,
the abundance is suppressed by $\sim (m_\mu/m_\tau)^2$ and given by
\bea
Y_a^{{\rm (th)}} &\sim&2 \times 10^{-5} \lrfp{g_*}{106.75}{-1} \lrf{T_R}{ \GEV{8}} \lrfp{f_a}{\GEV{10}}{-2}.
\eea
In this case successful thermal leptogenesis may be possible~\cite{Fukugita:1986hr}, with a mild degeneracy among
the right-handed neutrinos.
 Note that the thermally produced
PNGBs of $7$keV mass behave as warm dark matter because of their non-negligible free streaming.

The PNGBs can also be produced by the initial misalignment mechanism. The PNGB starts to oscillate when 
the Hubble parameter becomes comparable to the mass $m_a$. In the radiation dominated Universe, 
this happens when $T \sim 2 \times \GEV{6}  (m_a/7{\rm keV})^{1/2}$. Therefore, for $T_R \lesssim \GEV{6}$, 
the oscillations starts before the reheating, and the PNGB abundance is given by
\bea
Y_a^{\rm (mis)} &\sim& 3 \times 10^{-7} \lrf{T_R}{\GEV{6}} \lrfp{m_a}{7{\rm keV}}{-1} \lrfp{f_a}{\GEV{10}}{2} \theta_*^2,
\eea
where $\theta_* \equiv a_{\rm ini}/f_a$ denotes the initial oscillation amplitude. If the U(1)$_F$  symmetry is spontaneously broken
after inflation, we should replace $\theta_*$ with its averaged value, $\sqrt{\langle \theta_*^2\rangle} = \pi/\sqrt{3}$.\footnote{
Recently, the BICEP2 experiment found the primordial B-mode polarization, implying that the inflation scale is about $H_{\rm inf}
\sim \GEV{14}$~\cite{BICEP2}. If this is true, the global U(1)$_F$  symmetry must become spontaneously broken after inflation 
to avoid generating too large isocurvature perturbations. In this case, one needs to introduce extra breaking terms
to avoid the cosmological catastrophe induced by domain walls. 
}
For $T_R \gtrsim 2 \times \GEV{6}$, the abundance of PNGBs produced by the initial misalignment mechanism becomes
independent of $T_R$.
Therefore,  the initial misalignment mechanism is subdominant compared to the thermal production for $f_a = \GEV{10}$.
Note that the dependence of the abundance on $f_a$ is different between the two production processes,
and that for slightly larger values of $f_a$, the initial misalignment mechanism can dominate over the thermal production.
This is the case if $q_e$ is comparable to $\sim 3$ or larger.

Lastly we consider the case of $q_e = 0$. In this case the preferred value of $f_a$ is about $\GEV{5}$, and the thermal
production always dominate over the initial misalignment mechanism unless the anharmonic effect becomes significant~\cite{Turner:1985si,Lyth:1991ub,Visinelli:2009zm,Kobayashi:2013nva}.
  For $T_R$ above the weak scale,
the PNGBs are thermalized.
For $m_\mu < T < m_\tau$,   the PNGBs can be produced by scattering processes such as
$\mu + \gamma \to \mu + a$ with a rate given by
\bea
\Gamma_{\mu + \gamma \to \mu + a} \sim \alpha_{em} \frac{m_\mu^2}{f_a^2} T,
\eea
where $T$ is the temperature. The production
through the above process
is most efficient at $T = m_\mu$, and the production rate exceeds the Hubble parameter 
at that time if
\bea
f_a &\lesssim& 4 \times \GEV{7}.
\eea
Therefore, for $T_R \gtrsim m_\mu$, the PNGBs are thermalized,  and we need an additional entropy dilution by a factor of $40$.\footnote{
If the PNGB mass is of  ${\cal O}(0.1)$\,eV or lighter, there is no problem even if it is thermalized.
It would contribute to hot dark matter~\cite{Archidiacono:2013cha,Jeong:2013oza} or the effective neutrino species 
$\Delta N_{\rm eff} \simeq  0.39$~\cite{Weinberg:2013kea}. Their existence  are favored by 
recent observations~\cite{Wyman:2013lza,Hamann:2013iba,Battye:2013xqa,Ade:2013zuv}.
Interestingly, hot dark matter or dark radiation can relax the tension between BICEP2 and Planck.
}
If $T_R = {\cal O}(10)$ MeV, it is possible to produce the right amount of PNGBs to account for the observed dark matter abundance.

\section{Anomaly-free flavor model for leptons}
\label{sec:4}
In this section we build anomaly-free flavor models for leptons.
For simplicity we focus on a case in which electrons and muons are charged 
under the U(1)$_F$  symmetry, while tau leptons are neutral. The extension to a more general
charge assignment is straightforward. 

\begin{table}[t]
\begin{center}
\begin{tabular}{c||c|c|c|c|c|c}
 &$e_R$&$\mu_R$&$\tau_R$&$\ell_1$&$\ell_2$&$\ell_3$\\ \hline
 $Q$&$-a$&$-b$&0&$c$&$d$&0\\
\end{tabular}
\end{center}
\caption{The charge assignment of leptons under the global U(1)$_F$  flavor symmetry.}
\label{Q}
\end{table}%

Let us parametrize the global U(1)$_F$  charges of $e_{i}$ and $\ell_j$
as $Q(e_i)=(-a ,-b,0)$ and $Q(\ell_j)=(c,d,0)$, where $e_{i}$ and $\ell_j$ are 
 the right-handed charged-lepton singlet and the left-handed lepton doublet,
respectively, and  the subindices $i,j = 1,2,3$ represent the generation.
The charge assignment is also shown in  Table~\ref{Q}. 
As long as there are no other fermions
charged under both  the global U(1)$_F$ and SM gauge symmetries, the absence
of the SM gauge anomalies  requires
\bea
&&a +b = 0 \\
&&c+d =0.
\eeq

In order to write down Yukawa interactions for leptons, 
we need Higgs fields charged under the U(1)$_F$  symmetry.
Although not mandatory, let us seek the charge assignment, for which the off-diagonal 
elements are forbidden by the U(1)$_F$  symmetry.  We introduce three Higgs doublets,
$H(0)$,  $H(a+c)$, and  $H(-a-c)$, and require  the following conditions:
\bea
\label{cond1}
&&a \ne 0 \\
\label{cond2}
&& c \ne 0\\
\label{cond3}
&& a \ne c\\
\label{cond4}
&&2a+c \ne 0\\
\label{cond5}
&&a+2c \ne 0
\eea
For any charged assignment satisfying the above conditions, the  Yukawa
interactions take the diagonal form,
\beq
{\cal L} \supset
y_e {\bar e}_R \ell_1 H(-a-c)+
y_\mu {\bar \mu}_{R} \ell_2 H(a+c)
+ y_\tau {\bar \tau}_{R} \ell_3 H(0)+{\rm h.c.}
\eeq
Let us normalize the global U(1)$_F$  charge so that  $c = 1$.  Then the above conditions from (\ref{cond1}) 
to (\ref{cond5})  read
$a \ne 0, 1,-2$, and so, the allowed integer values of $a$ are $a=-1, +2, \pm3, \cdots$.
Let us take up the first two cases, namely, $a=-1$ and $a=2$.

\subsection{Case of $a=-1$}
In this case the global U(1)$_F$  symmetry is identical to $L_e - L_\mu$,
and there is only one Higgs doublet, $H(0)$. 
Then the PNGB does not have direct couplings with charged leptons like \EQ{NGlepton}, as
it does not reside in the phase of $H(0)$.

Let us  extend the SM to include right-handed  neutrinos $N_i$.  
If  we assign the global U(1)$_F$  charges as  $Q(N_i) = (1,-1,0)$,
the neutrino Yukawa interaction is diagonal;
\beq
{\cal L}\;\supset\; 
y_{i}^N {\bar N}_i
\ell_i {\tilde H}(0) + {\rm h.c.}
\eeq
where ${\tilde H}(0) = i \sigma_2 H(0)^*$.
 The observed large neutrino mixings can be explained if the 
Majorana mass matrix for $N_i$ contains large off-diagonal elements. 
To this end we introduce U(1)$_{B-L}$ gauge symmetry and the 
$B-L$ Higgs fields, $\phi(0)$, $\phi(\pm1)$, and $\phi(\pm2)$,
where  the numbers in the parentheses represent  the U(1)$_F$  charge and 
 they are assumed to have a common $B-L$ charge $+2$. 
 If these $B-L$ Higgs fields develop
non-zero vacuum expectation values (VEVs),  the U(1)$_F$  symmetry is spontaneously 
 broken, and the Majorana  mass matrix for $N_i$ is induced as
\bea
-{\cal L} \supset \frac{1}{2} (M_{N})_{ij} {\bar N}^c_i N_j + {\rm h.c.}
\eea
with
\beq
M_N \sim
\left(
\bear{ccc}
\phi(-2)&\phi(0)&\phi(-1)\\
\phi(0)&\phi(2)&\phi(1)\\
\phi(-1)&\phi(1)&\phi(0)
\eear
\right),
\eeq
where the $B-L$ Higgs fields are understood to represent their VEVs,
and we have dropped $O(1)$ numerical coefficient in 
each element.  If the VEVs  are comparable to each other,  the large neutrino mixing angles are realized. 
The light neutrino masses can be explained by the seesaw mechanism~\cite{seesaw}.

The PNGB resides in the phase of $\phi(1)$ and $\phi(2)$, and the decay constant $f_a$ is
approximately given by their VEVs.  In fact, the PNGB in this case is similar to the majoron.
The cosmological constraints on the majoron dark matter were studied in e.g. Ref.~\cite{Lattanzi:2013uza}.

One can also introduce  the Higgs portal couplings $\sim |\phi|^2 |H|^2$.  The situation would be similar to the model proposed by Weinberg~\cite{Weinberg:2013kea}. 
For a certain set of parameters, massless PNGBs would contribute the effective neutrino species, $\Delta N_{\rm eff}$.

\subsection{Case of $a=+2$}
In this case the charge assignment is $Q(e_{i}) = (-2,2,0)$ and $Q(\ell_i)=(1,-1,0)$,
and there are three Higgs doublets, $H(0)$ and $H(\pm3)$. 
The charged lepton Yukawa interactions are given by
\beq
{\cal L}\;\supset\; 
y_e {\bar e}_R \ell_1 H(-3)
+y_2 {\bar \mu}_R \ell_2 H(3)
+y_3 {\bar \tau}_R \ell_3 H(0)
+ {\rm h.c.}.
\eeq
Th previous argument on the neutrino Yukawa interaction and the right-handed neutrino mass matrix
 can be applied to the present case, and the observed large neutrino mixing
as well the neutrino mass scale can be similarly explained.

The global U(1)$_F$  symmetry is spontaneously broken by both the Higgs doublets and the $B-L$ Higgs fields. 
We assume that the symmetry breaking scale is of order $\GEV{10}$ (or smaller).
 Then, while the PNGB resides mainly in the phase of $\phi(1)$ and $\phi(2)$, 
 it also appears in the phase of $H(3)$ and $H(-3)$, and so,  electrons and muons are coupled to the PNGB in the low
 energy as in \EQ{NGlepton}.
Since the PNGB does not have (sizable) couplings with gluons, photons, and quarks,
the astrophysical bounds  are considered to be rather weak.

A couple of comments are in order. In order to give a mass to the PNGB, one needs an explicit  U(1)$_F$ 
symmetry breaking. It is interesting to note that the following term
\bea
{\cal L}_{breaking} &=& m^2 H(-3)^\dag H(3) + {\rm h.c.}
\eea
breaks the U(1)$_F$  symmetry down to the  subgroup $Z_6$, giving rise to a PNGB mass $m_a \sim {\cal O}(1)$\,keV 
for $m \sim \langle H(-3) \rangle \sim \langle H(3) \rangle \sim \GEV{2}$   and $f_a = {\cal O}(10^{10})$\,GeV. 
Therefore, the PNGB associated with anomaly-free flavor symmetry broken at $f_a = {\cal O}(10^{10})$\,GeV
nicely explains both the mass and the lifetime suggested by the observed $3.5$\,keV X-ray line.

The off-diagonal elements of  the charged lepton Yukawa matrix receive non-zero contributions, 
as the U(1)$_F$  symmetry is spontaneously broken. 
Their contributions to the lepton-flavor violating processes, however,  are negligible in our model.

In the presence of the $B-L$ Higgs fields, there are in general
mixings between  $H$ and $\phi$. Such mixings are assumed to be small in our context to keep the hierarchy
between the weak scale and the flavor symmetry breaking scale.
Also we assume that the lightest Higgs has a property similar to the SM Higgs and the other Higgs fields
are so heavy that they evade the current collider search. Some of them, however, may be
within the reach of LHC and/or ILC.

\section{Discussion and conclusions}
\label{sec:5}

Some comments and discussions are in order. In the case of $q_e \ne 0$, 
$f_a = {\cal O}(10^{9-10})$\,GeV is needed to explain the $3.5$\,keV X-ray line. 
Since the couplings to photons, gluons and nucleons are suppressed, 
the PNGBs avoid various astrophysical and ground-based constraints.
Still, it may be possible to find them in the future. 
Interestingly,  there is a hint for an extra cooling of white dwarfs, which can be
explained  by light PNGBs coupled to electrons with the decay constant in this range~\cite{Isern:1992gia}.\footnote{
In Ref.~\cite{Isern:1992gia}, the QCD axion was considered,
and so, the cooling rate due to the $7$\,keV axion can be much smaller for the same decay rate. 
There may be another PNGB, if the flavor symmetry group larger than U(1)$_F$.
} 
If  such light PNGBs are coupled with electrons but not with photons, it is possible that they are copiously produced
in the Sun, but cannot be detected by experiments using the magnetic field like the CAST experiment~\cite{Barth:2013sma}.

In the case of $q_e = 0$ and $q_\mu \ne 0$, the preferred value of $f_a$ is of order $\GEV{5}$, much smaller than
the previous case. Still, as the effective PNGB coupling to the photon is so weak that the constraint from the cooling 
of horizontal branch stars can be satisfied~\cite{Raffelt:1996wa}.
On the other hand, the bound from supernova cooling will be more non-trivial since
the PNGB couples to muons directly and the muons might be abundant in the supernova core~\cite{Canuto:1975kr,Colucci:2013pya}.
Although the muon abundance depends sensitively on the temperature, the preferred value of $f_a$ may be in tension
with the observation. 
As a rough estimate, we refer to the constraint on the majoron coupling constants to neutrinos from the supernova cooling:
it is bounded as $g_{ee} \lesssim 10^{-6}$ where $g_{ee}$ is the yukawa coupling between the majoron and electron neutrinos~\cite{Farzan:2002wx}. In our case, the effective coupling constant between the PNGB and the muon reads $m_{\mu}/f_a \sim 10^{-6}$. A more
detailed study is needed to test the viability of this model.

So far we have considered the U(1)$_F$  flavor symmetry, under which only leptons are charged,
and we constructed models in which the lepton mass matrix is (almost) diagonal. It is possible
to extend the models to allow larger off-diagonal terms,  or to extend the flavor symmetry to
the quark sector, by enlarging the flavor symmetry and adding more Higgs fields.
If the actual flavor symmetry group is larger than U(1)$_F$  
and if it is broken at a scale of ${\cal O}(10^{9-10})$\,GeV, there may be more PNGBs with different masses with or without
couplings to photons and/or gluons. 
Then it may be possible to provide a unified picture of the QCD axion well as other PNGBs.
In this case the light PNGBs can be searched for by flavor-changing processes such as
$\tau \to \mu + a$, $\mu^+ \to e^+ + a$, $K^+ \to \pi^+ + a$~\cite{Gelmini:1982zz}.

We have pursued a possibility that  a PNGB is lurking below the intermediate scale,
evading the astrophysical bounds. Along this lines we have proposed flavor models based on
an anomaly-free U(1)$_F$  symmetry,  where the PNGB is preferentially coupled to the leptons.
In particular, its anomalous couplings to gluons and photons are  absent, greatly relaxing the astrophysical bounds. We have also pointed
out that, although suppressed, the PNGB coupling to photons is induced by threshold 
corrections. Interestingly, the recent hint for the X-ray line at about $3.5$\,keV~\cite{Bulbul:2014sua,Boyarsky:2014jta} 
can be explained by the PNGB dark matter with $m_a \simeq 7$\,keV for the decay constant
$f_a = \GEV{9-10}$ ($f_a = \GEV{5-6}$) if electrons are (not) charged under the flavor symmetry.

\section*{Acknowledgments}
This work was supported by the Grant-in-Aid for Scientific Research on
Innovative Areas (No. 21111006  [KN and FT],  No.23104008 [FT], No.24111702 [FT]),
Scientific Research (A) (No. 22244030 [KN and FT], 21244033 [FT], 22244021 [TTY]),  JSPS Grant-in-Aid for
Young Scientists (B) (No.24740135) [FT], and Inoue Foundation for Science [FT].  This work was also
supported by World Premier International Center Initiative (WPI Program), MEXT, Japan.


\begin{thebibliography}{99}

\bibitem{Peccei:1977hh}
  R.~D.~Peccei and H.~R.~Quinn,
  Phys.\ Rev.\ Lett.\  {\bf 38}, 1440 (1977);
  Phys.\ Rev.\ D {\bf 16}, 1791 (1977).


\bibitem{QCD-axion}
  For a review, see
  J.~E.~Kim,
  Phys.\ Rept.\  {\bf 150}, 1 (1987);
  H.~Y.~Cheng,
  Phys.\ Rept.\  {\bf 158}, 1 (1988);
  J.~E.~Kim and G.~Carosi,
  Rev.\ Mod.\ Phys.\ \ {\bf 82}, 557  (2010);
 A.~Ringwald,
 Phys.\ Dark Univ.\  {\bf 1} (2012) 116;
  M.~Kawasaki and K.~Nakayama,
  Ann.\ Rev.\ Nucl.\ Part.\ Sci.\  {\bf 63}, 69 (2013)
  [arXiv:1301.1123 [hep-ph]].


\bibitem{Jaeckel:2010ni} 
See e.g. 
  J.~Jaeckel and A.~Ringwald,
  Ann.\ Rev.\ Nucl.\ Part.\ Sci.\  {\bf 60}, 405 (2010)
  [arXiv:1002.0329 [hep-ph]];
  P.~Arias, D.~Cadamuro, M.~Goodsell, J.~Jaeckel, J.~Redondo and A.~Ringwald,
  JCAP {\bf 1206}, 013 (2012)
  [arXiv:1201.5902 [hep-ph]].
  

  
\bibitem{Andreas:2010ms} 
  S.~Andreas, O.~Lebedev, S.~Ramos-Sanchez and A.~Ringwald,
  JHEP {\bf 1008}, 003 (2010)
  [arXiv:1005.3978 [hep-ph]].
  
\bibitem{Cadamuro:2011fd} 
  D.~Cadamuro and J.~Redondo,
  JCAP {\bf 1202}, 032 (2012)
  [arXiv:1110.2895 [hep-ph]].
  

  
\bibitem{Weinberg:2013kea} 
  S.~Weinberg,
  Phys.\ Rev.\ Lett.\  {\bf 110}, no. 24, 241301 (2013)
  [arXiv:1305.1971 [astro-ph.CO]].
  
\bibitem{Essig:2010gu} 
  R.~Essig, R.~Harnik, J.~Kaplan and N.~Toro,
  Phys.\ Rev.\ D {\bf 82}, 113008 (2010)
  [arXiv:1008.0636 [hep-ph]].
  

\bibitem{Bulbul:2014sua}
  E.~Bulbul, M.~Markevitch, A.~Foster, R.~K.~Smith, M.~Loewenstein and S.~W.~Randall,
  arXiv:1402.2301 [astro-ph.CO].

\bibitem{Boyarsky:2014jta}
  A.~Boyarsky, O.~Ruchayskiy, D.~Iakubovskyi and J.~Franse,
  arXiv:1402.4119 [astro-ph.CO].
  
\bibitem{Higaki:2014zua} 
  T.~Higaki, K.~S.~Jeong and F.~Takahashi,
  arXiv:1402.6965 [hep-ph].
  
\bibitem{Kawasaki:1997ah} 
  M.~Kawasaki and T.~Yanagida,
  Phys.\ Lett.\ B {\bf 399}, 45 (1997)
  [hep-ph/9701346].

\bibitem{Hashiba:1997rp}
  J.~Hashiba, M.~Kawasaki and T.~Yanagida,
  Phys.\ Rev.\ Lett.\  {\bf 79}, 4525 (1997)
  [hep-ph/9708226];
  T.~Asaka, J.~Hashiba, M.~Kawasaki and T.~Yanagida,
  Phys.\ Rev.\ D {\bf 58}, 083509 (1998)
  [hep-ph/9711501];
  Phys.\ Rev.\ D {\bf 58}, 023507 (1998)
  [hep-ph/9802271].
  
\bibitem{Kusenko:2012ch} 
  A.~Kusenko, M.~Loewenstein and T.~T.~Yanagida,
  Phys.\ Rev.\ D {\bf 87}, no. 4, 043508 (2013)
  [arXiv:1209.6403 [hep-ph]].
  
    

\bibitem{Ishida:2014dlp} 
  H.~Ishida, K.~S.~Jeong and F.~Takahashi,
  arXiv:1402.5837 [hep-ph];
  see also 
id., 
  arXiv:1309.3069 [hep-ph], to appear in Phys. Lett. B.
  
\bibitem{Finkbeiner:2014sja} 
  D.~P.~Finkbeiner and N.~Weiner,
  arXiv:1402.6671 [hep-ph].
  


  
\bibitem{Jaeckel:2014qea} 
  J.~Jaeckel, J.~Redondo and A.~Ringwald,
  arXiv:1402.7335 [hep-ph].
  
\bibitem{Lee:2014xua} 
  H.~M.~Lee, S.~C.~Park and W.~-I.~Park,
  arXiv:1403.0865 [astro-ph.CO].
  

  
\bibitem{Krall:2014dba} 
  R.~Krall, M.~Reece and T.~Roxlo,
  arXiv:1403.1240 [hep-ph].
  
  
\bibitem{Jaeckel:2013uva} 
  J.~Jaeckel,
  arXiv:1311.0880 [hep-ph].
  
\bibitem{Fukugita:1986hr} 
  M.~Fukugita and T.~Yanagida,
  Phys.\ Lett.\ B {\bf 174}, 45 (1986).
  
    \bibitem{BICEP2}
  P.~A.~RAde {\it et al.}  [ BICEP2 Collaboration],
  arXiv:1403.3985 [astro-ph.CO].
  
  
\bibitem{Turner:1985si}
  M.~S.~Turner,
  Phys.\ Rev.\ D {\bf 33}, 889 (1986).

\bibitem{Lyth:1991ub}
  D.~H.~Lyth,
  Phys.\ Rev.\ D {\bf 45}, 3394 (1992).


\bibitem{Visinelli:2009zm}
  L.~Visinelli, P.~Gondolo and ,
  Phys.\ Rev.\ D {\bf 80}, 035024 (2009)
  [arXiv:0903.4377 [astro-ph.CO]].


\bibitem{Kobayashi:2013nva}
  T.~Kobayashi, R.~Kurematsu and F.~Takahashi,
  JCAP {\bf 1309}, 032 (2013)
  [arXiv:1304.0922 [hep-ph]].

  
\bibitem{Archidiacono:2013cha} 
  M.~Archidiacono, S.~Hannestad, A.~Mirizzi, G.~Raffelt and Y.~Y.~Y.~Wong,
  JCAP {\bf 1310}, 020 (2013)
  [arXiv:1307.0615 [astro-ph.CO]].
  
\bibitem{Jeong:2013oza} 
  K.~S.~Jeong, M.~Kawasaki and F.~Takahashi,
  JCAP {\bf 1402}, 046 (2014)
  [arXiv:1310.1774 [hep-ph], arXiv:1310.1774].
  
\bibitem{Wyman:2013lza}
  M.~Wyman, D.~H.~Rudd, R.~A.~Vanderveld and W.~Hu,
  Phys.\ Rev.\ Lett.\  {\bf 112}, 051302 (2014)
  [arXiv:1307.7715 [astro-ph.CO]].

\bibitem{Hamann:2013iba}
  J.~Hamann and J.~Hasenkamp,
  JCAP {\bf 1310}, 044 (2013)
  [arXiv:1308.3255 [astro-ph.CO]].

\bibitem{Battye:2013xqa}
  R.~A.~Battye and A.~Moss,
  Phys.\ Rev.\ Lett.\  {\bf 112}, 051303 (2014)
  [arXiv:1308.5870 [astro-ph.CO]].

  \bibitem{Ade:2013zuv}   P.~A.~R.~Ade {\it et al.}  [Planck Collaboration],  
  arXiv:1303.5076 [astro-ph.CO].

       \bibitem{seesaw}
T.~Yanagida, in Proceedings of the {\it{``Workshop on the Unified Theory and
 the Baryon Number in the Universe''}}, Tsukuba, Japan, Feb. 13-14, 1979, edited by
O.~Sawada and A.~Sugamoto, KEK report KEK-79-18, p. 95,
and {\it{``Horizontal Symmetry And Masses Of Neutrinos''
}}, Prog. Theor. Phys. {\bf{64}} (1980) 1103;
M.~Gell-Mann, P.~Ramond and R.~Slansky, in {\it{``Supergravity''}}
 (North-Holland, Amsterdam, 1979) {\it{eds}}. D.~Z.~Freedom and P.~van
Nieuwenhuizen, Print-80-0576 (CERN);
see also P.~Minkowski,  Phys.\ Lett.\  B {\bf 67}, 421 (1977).

  
\bibitem{Lattanzi:2013uza} 
  M.~Lattanzi, S.~Riemer-Sorensen, M.~Tortola and J.~W.~F.~Valle,
  Phys.\ Rev.\ D {\bf 88}, 063528 (2013)
  [arXiv:1303.4685 [astro-ph.HE]].
  


  

    

    
\bibitem{Isern:1992gia} 
  J.~Isern, M.~Hernanz and E.~Garcia-Berro,
  Astrophys.\ J.\  {\bf 392}, L23 (1992);
  J.~Isern, L.~Althaus, S.~Catalan, A.~Corsico, E.~Garcia-Berro, M.~Salaris, S.~Torres and L.~Althaus {\it et al.},
  arXiv:1204.3565 [astro-ph.SR].
  
\bibitem{Barth:2013sma} 
  K.~Barth, A.~Belov, B.~Beltran, H.~Brauninger, J.~M.~Carmona, J.~I.~Collar, T.~Dafni and M.~Davenport {\it et al.},
  JCAP {\bf 1305}, 010 (2013)
  [arXiv:1302.6283 [astro-ph.SR]].
  
\bibitem{Raffelt:1996wa} 
  G.~G.~Raffelt,
  ``Stars as laboratories for fundamental physics : The astrophysics of neutrinos, axions, and other weakly interacting particles,''
  Chicago, USA: Univ. Pr. (1996) 664 p
  
  
\bibitem{Canuto:1975kr} 
  V.~Canuto,
  Ann.\ Rev.\ Astron.\ Astrophys.\  {\bf 13}, 335 (1975).
  
\bibitem{Colucci:2013pya} 
  G.~Colucci and A.~Sedrakian,
  Phys.\ Rev.\ C {\bf 87}, 055806 (2013)
  [arXiv:1302.6925 [nucl-th]].
  
\bibitem{Farzan:2002wx} 
  Y.~Farzan,
  Phys.\ Rev.\ D {\bf 67}, 073015 (2003)
  [hep-ph/0211375].
  
\bibitem{Gelmini:1982zz} 
  G.~B.~Gelmini, S.~Nussinov and T.~Yanagida,
  Nucl.\ Phys.\ B {\bf 219}, 31 (1983).
  
\end{thebibliography}
\end{document}